# Exploring Angle dependent Phonon Modes in Sodium Mesitylene Sulfonate (SMS) crystals using THz-time domain polarimetry (THz-TDP)

Yamuna Murtunge,*,†,‡ Ajinkya Punjal,†,¶ Ruturaj Puranik,† Utkarsh Pandey,† S.B. Kulkarni,‡ and S.S. Prabhu*,†

†Department of Condensed Matter Physics and Material Science, Tata Institute of Fundamental Research, Mumbai
‡Department of Physics, Dr Homi Bhabha State University, Institute of Science, Mumbai
¶Department of Physics, University of Cambridge, U.K

E-mail: murtungeyamuna31@gmail.com ; prabhu@tifr.res.in

**Abstract:**
We employed sodium mesitylene sulfonate crystals to investigate angle-dependent phonon resonance and thickness dependent splitting in THz time-domain polarimetry. This crystal possesses a C2 space group, leading to a repetition pattern after 180° rotations. Our experimental observations revealed intriguing behaviour: We observed a non-linear response when varying the angle from 10° to 360° in both 0.182mm and 1.266mm thick crystals. specifically, at 90° and 270°, dip resonance occurred, while at 180° and 360°, no phonon resonance was observed. For thick crystal we observed the splitting of phonon modes. Our findings offer valuable insights into the phononic properties of this crystal as the angle varies.

## 1 Introduction

Organic crystals have recently become a captivating material in the realm of Terahertz time-domain spectroscopy (THz-TDS) due to their ability to generate highly intense THz radiation and their applications in various THz devices, such as detectors and modulators [1]. But the growing response of organic crystals as sources in THz frequency needs to study the behaviour of crystals, particularly the phonon modes that arise from inter and intramolecular vibrations of the molecule because it crucially affects the bandwidth of THz generation, efficiency and capabilities [2].

The crystal's anisotropic nature allows us to explore angle-dependent phonon modes by manipulating the sample orientation and observing distinct changes in phonon mode intensities and polarisations. Understanding how phonon modes behave with varying angles in organic crystals is essential for unravelling their anisotropic functionalities.

This paper aims to investigate and characterise the angle dependent occurrence of phonon resonance and thickness dependent splitting in sodium mesitylene sulfonate (SMS) single crystal using THz time-domain polarimetry (THz-TDP). We focus specifically on the angle-dependent behaviour of the refractive index, dielectric constant, and conductivity. The THz time domain polarimetry technique is utilised to gain insights into this crystal's vibrational behaviour, phononic properties, and polarisation effects. Determining the polarization state of the transmitted THz wave through the crystal will give information about the various modes present in the crystal for a particular orientation. It will lead to a better understanding of the crystal properties.

## 2 Results and discussion:

We successfully grew the nonlinear SMS crystal with dimensions of 10x5x0.182mm3, 0x5x0.35mm3 and 10x5x1.266mm3 using a slow evaporation method with methanol as a solvent [3]. According to the XRD data, this crystal belongs to the monoclinic crystal structure with a non-centrosymmetric C2 space group, and the lattice parameters are a=8.6926Å, b=7.3679Å, c=16.4519Å, and α=γ=90◦, β=103.7790◦

The THz transmission and the state of polarization of sodium mesitylene sulfonate crystal are assessed using Femtosecond pulsed laser FemtoSource Synergy (Femtolasers GmbH), generating 10 fs pulses at an 80 MHz repetition rate with a centre wavelength of 800nm. The setup includes four parabolic mirror configurations. The laser beam is divided into the pump beam, which is utilised to generate THz radiation employing an LT-GaAs-based photoconductive antenna, and the probe beam, which is used to detect the THz signal. The resulting THz signal is seen with the help of a 2 mm thick ZnTe < 110 > orientation using a standard electro-optic method. The entire setup is confined within a nitrogen-purged box to eliminate any unwanted absorption of water lines in the air. [4]

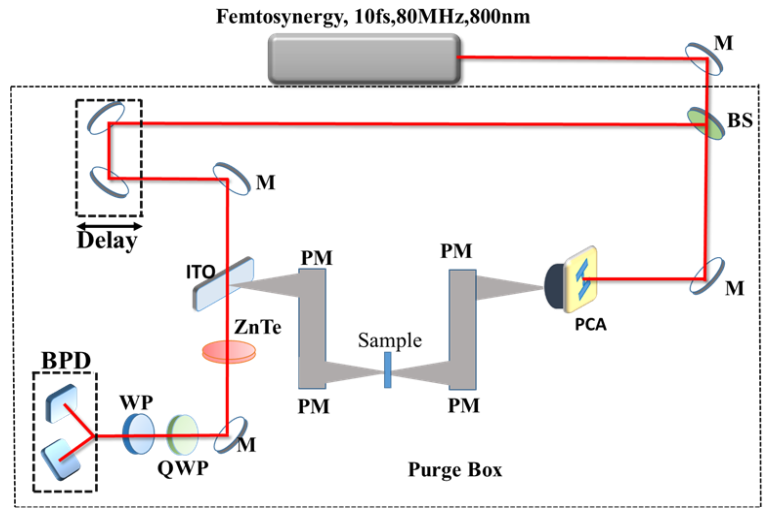

Figure 1: THz-TDS set-up used to characterise the SMS crystal
M: Mirror, BS: Beam splitter, PCA: Photoconductive antenna, PM: Parabolic Mirror, ITO: Indium Tin Oxide, ZnTe: Zinc Telluride, QWP: Quarter Wave Plate, WP: Wollaston Prism, BPD: Balanced Photo-Detectors

## 2.1 Polarisation Dependent THz-TDS Measurements:

To gather the transmission spectra, we demonstrated THz time-domain spectroscopy (THz TDS) and analysed the polarisation state of a sodium mesitylene sulfonate crystal (001). We used 0.182mm, 0.35mm and 1.266mm thick crystals perpendicular to the incident electric field. On the same crystals, we performed rotations of the sample within the *xy* plane and observed a significant change in absorption over a 10° interval. Notably, 0.182mm thick crystal displayed a non-linear response. Crystal exhibited dip resonances at two distinct frequencies (0.94THz and 1.922THz) when varying the angle from 10° to 360°, particularly prominent at 90° and 270° while at 180° and 360° no phonon modes were present. Furthermore, we investigated the crystal's vibrational behaviour as the polarisation angle increased from 10° to 360°.

The THz frequency domain transmission was obtained by measuring the signal response without and with samples in the time domain and then dividing the Fast Fourier-transformed reference (Air scan) signal by the sample shown in Fig.2a and Fig.2b. The results underscore the significant influence of crystallographic orientation on the lattice dynamics and phonon interaction within the crystal.

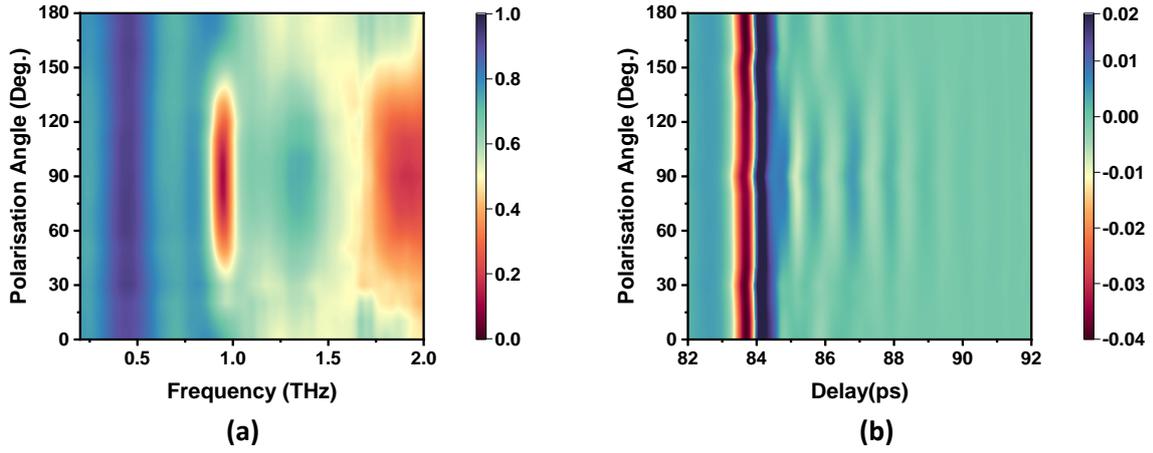

Figure 2: (a)Transmittance THz spectrum of the SMS crystal at room temperature. Transmission shows minima at two angles for specific frequencies. (b) THz Pulse transmission through the SMS crystal shows different oscillations at specific angles.

We use terahertz time-domain polarimetry to extract the angle-dependent complex refractive index, complex dielectric constant, and complex conductivity in the frequency range of 0.2 to 2.5THz. The absorption peaks around 0.94THz and 1.922THz could be due to the strong interaction between the incident electric field and the electric dipole moment associated with the vibrating atoms in the lattice of the molecule. These observations highlight the crystals highly anisotropic. Since this crystal has C2 symmetry, data reproduced over 180°-360°, attributed to its C2 space group [5]. When incident THz radiation is coupled with crystals' phonon modes, a strong interaction occurs between the lattice vibration of the crystal and the THz waves, leading to a change in molecular properties. If both frequencies match, the oscillating lattice polarization causes an increase in the refractive index shown in Fig.3a and Fig.3b SMS crystal as a function of frequency. The index of refraction at absorbing modes 0.94 THz ranges from ~1.65 to ~2.25, while for 1.92 THz, it is from ~1.95 to ~2.15, respectively.

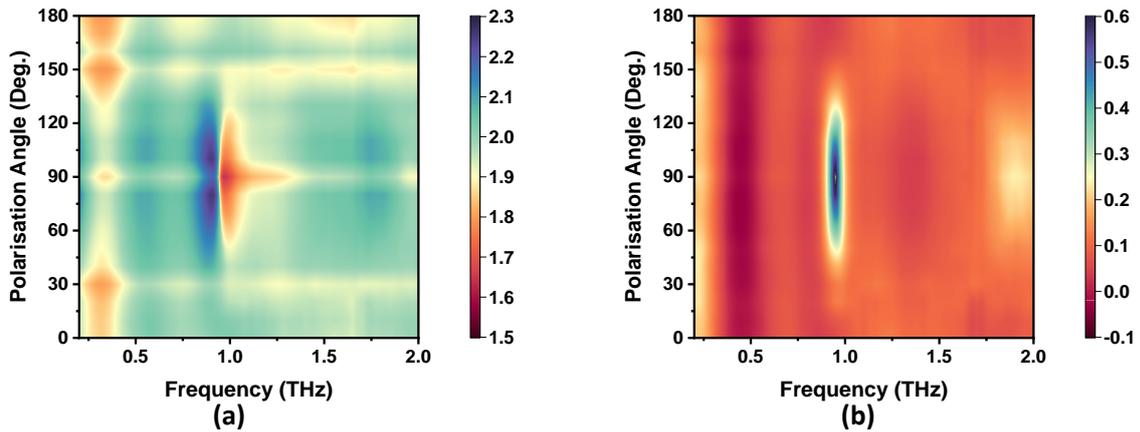

Figure 3: (a)Numerically Extracted Real part of the Refractive index shows absorption feature at a specific frequency and (b) the Extinction coefficient also shows the large value at the above-mentioned frequencies.

The complex dielectric constant $\tilde{\varepsilon} = \varepsilon_1 + i\varepsilon_2$ where $\varepsilon_1 = [n(\omega)]^2 - [\kappa(\omega)]^2$ and $\varepsilon_2 = [2n(\omega)\kappa(\omega)]$ [6] are determined from the real and imaginary parts of the refractive index given as,

$$\tilde{n} = n - ik$$

$$n = \left(\frac{c\varphi(\omega)}{\omega d}\right) + 1$$

$$k = \frac{c}{\omega d} \cdot \ln\frac{4n}{\rho(\omega)(n+1)^2}$$

In Fig.4a and Fig.4b, The strength of the electric dipole oscillator exhibit polarization dependent behaviour, is amplified at the specific polarization angle, such as 90° and 270°, leading to a rise in dielectric constant while its minimum at 180° and 360° polarization angles. The dielectric constant shows enhanced values at possible phonon frequencies of SMS crystals due to the excitement of lattice vibration or intermolecular vibrational modes when incident THz frequency is coupled with the vibrational modes of the crystal.

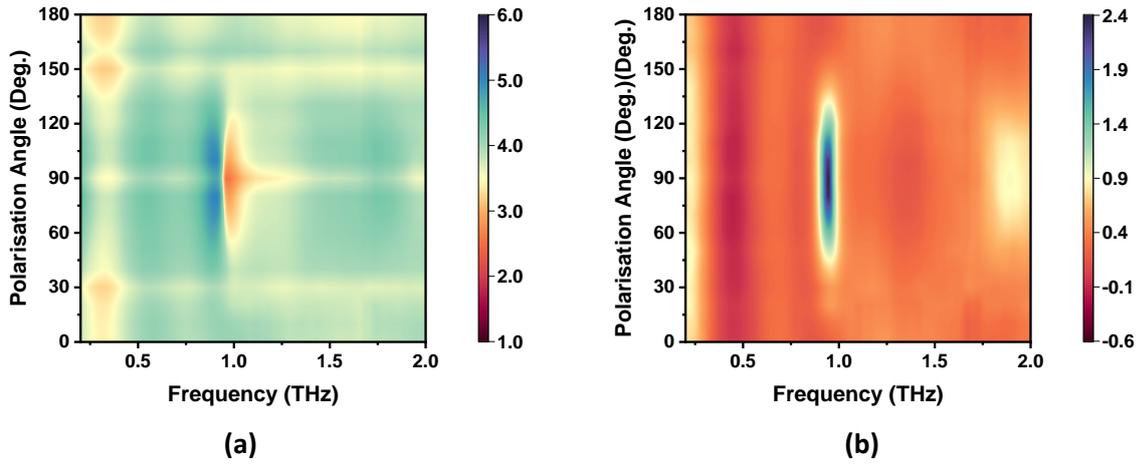

Figure 4: (a) Real part of the Dielectric constant and (b) Imaginary part of the Dielectric constant

Enhancing the dielectric constant at phonon modes can significantly impact the optical conductivity of the crystals. The complex conductivity in terms of a dielectric constant has been measured using the following equations:

$$\sigma' = \omega\varepsilon_o\varepsilon_2$$

$$\sigma'' = \omega\varepsilon_o(\varepsilon_\infty - \varepsilon_1)$$

where, $\varepsilon_o$ represents permittivity of free space and $\varepsilon_\infty$ signifies high-frequency dielectric constant, respectively.

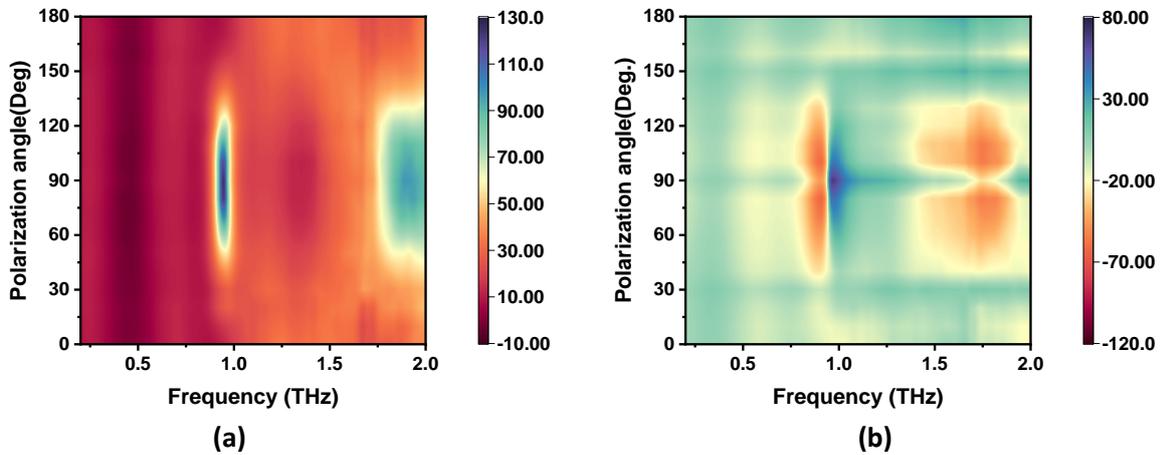

Figure 5: (a) Real and (b) Imaginary parts of the THz conductivity of SMS crystal at room temperature

As incident THz radiation interacts with the vibrational modes of the crystal, it induces phenomena such as phonon scattering and polaron formation. This phonon scattering typically reduces THz conductivity at resonances by impeding charge carrier mobility.

## 2.2 Thickness-Dependent THz-TDS Measurements:

The crystal's thickness plays a pivotal role in determining its phonon modes. Thicker crystals exhibit evolved phonon modes due to an extended interaction between incident THz radiation and the crystal lattice. In our study, as the crystal thickness increased to 1.266mm, we observed two split dips in the transmission spectra. These splits occurred at frequencies of approximately 0.8409 THz and 1.003 THz, shifting from the 0.945 THz observed in the 0.182mm crystal. The increase in crystal thickness also results in a higher number of allowed energy states for phonon resonance, leading to the splitting of phonon resonance into multiple branches, each corresponding to different quantised energy states. This thickness-dependent splitting provides a fascinating perspective on the vibrational properties of crystals and opens up possibilities for developing devices like THz detectors, modulators, and emitters. Our observation sheds light on a deeper understanding of the crystal lattice dynamics, possibly due to phonon-phonon interactions.

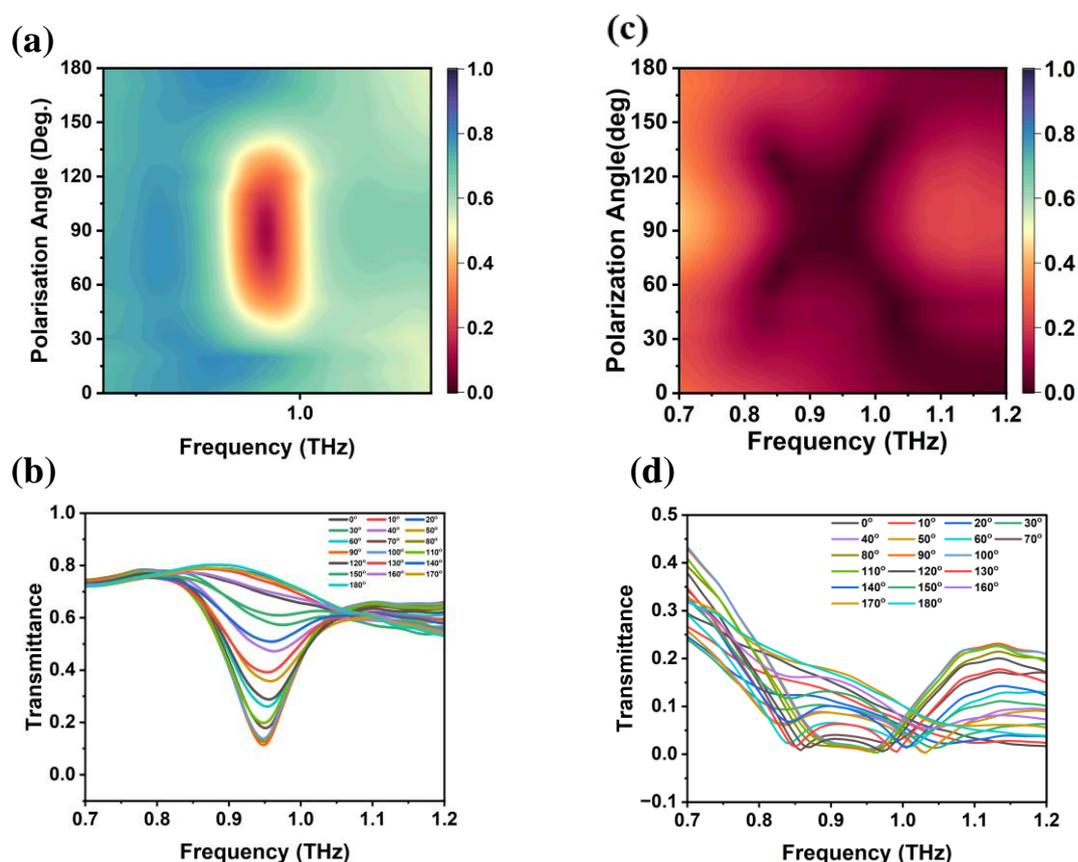

Figure 6: Thickness dependent splitting in organic SMS crystal: no coupling (a) & (b) for t=0.182mm, and Strong coupling(c)&(d) for 1.266mm thick

The sample stage is rotated from 0° to 180° during the measurement to obtain a transmission signal at different incident angles in the interval of 10°. The strong coupling of phonon-polariton observed in the range of 0.2–2.5 THz corresponds to the vibrational modes of the anion (sulphonate) and cation (Na) present in the molecular structure of the SMS crystal. Maximum splitting was observed when we rotate the sample from 40° to 120°. As this crystal is defective-free, we observed a substantial resonance dip at 0.94 THz for a 0.182 mm thick crystal; however, as thickness increased, we followed the appearance of two modes, upper polariton and lower polariton band and dispersed them with a variation of incident angle but original polariton band disappear. The strong coupling occurs when the interaction between the THz photons and crystals phonon vibration are sufficiently strong leading to formation of hybrid light matter states. This interaction results in Rabi splitting, where the energy levels split into two distinct branches observable as two separate peaks in the transmission spectra. The magnitude of the coupling strength is increasing due to better field confinement and volume as we increasing the thickness of the crystals shown in Fig. 6, the thickness-dependent transmittance spectra of 6(c) contour and 6(d) show significant curvature of the upper polariton band (UPB) and lower polariton band (LPB) formed due to a strong coupling effect, which indicates anti-crossing behaviour and evidence of Rabi splitting [7] between two modes. The multiple Lorentzian Oscillator model [9] for 90° of all crystals was applied to fit the experimentally obtained dielectric permittivity's of all three crystals given as,

$$\tilde{\varepsilon} = \varepsilon_\infty + \sum_{m=1}^{n} \frac{g_m}{(\omega_m^2 - \omega^2) + i\omega\gamma_m} = \varepsilon'(\omega) + i\varepsilon''(\omega)$$

where $\varepsilon_\infty$ = is the high-frequency limit of dielectric constant, $g_m$ (= $\Delta\varepsilon_m\omega^2_m$) represents the oscillator strength for the m$^{th}$ resonance, $\Delta\varepsilon_m$ is the dielectric strength, $\omega_m$ is the resonance frequency, and $\gamma_m$ is the damping factor of the resonance. The theoretical fitted parameters for all crystal thicknesses are shown in Table.1

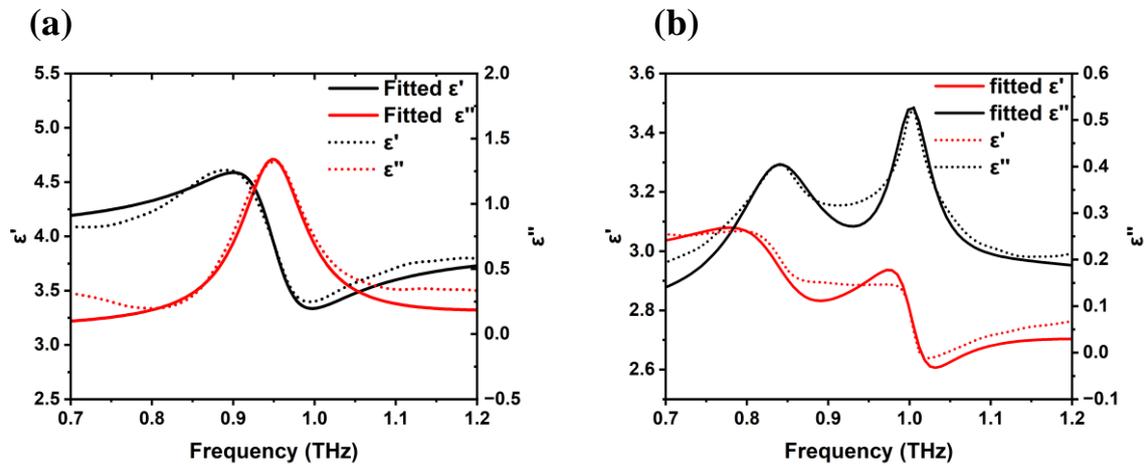

Figure 7: Experimentally obtained real ε '(Black solid line) and imaginary ε"(Red solid line) are shown with fitted(Dotted black and red lines) plots for (a)0.182mm (b)0.35mm and (c)1.266mm thick SMS crystal

| Crystal thickness | $\varepsilon_\infty$ | Resonant frequency $\omega_m$ (THz) | Oscillator strength $g_m$ | Scattering Time τ (ps) |
|---|---|---|---|---|
| t=0.182mm | 3.72 | 0.95 | 0.115±0.045 | 65±1.5 |
| t=1.266mm | 2.766 | 0.8409 | 0.03±0.025 | 49.5±1.2 |
| | | 1.003 | 0.0198±0.0023 | 108.5±3.6 |

Table:1 Dielectric parameters extracted using Lorentz multiple Oscillator model

## 3 Conclusion:

Coupling phonons with terahertz radiation substantially modifies the refractive index, dielectric properties, and electrical transport at phonon resonance frequencies. This enables selective control over THz waves in polar crystals. Our study highlights the angle-dependent occurrence of phonon modes in sodium mesitylene sulfonate crystal, as evidenced by THz time-domain polarimetry. We have demonstrated the thickness dependent enhancement in phonon coupling strength. These observed phenomena emphasise the significant influence of crystal orientation on the behaviour of phonon modes in the crystal. The second harmonic response demonstrates the SHG nature of the crystal.

## Acknowledgement:
The authors thank the Tata Institute of Fundamental Research, Mumbai, for providing instrumental facilities and resources that enable us to conduct this research successfully. We also wish to thank Mr Gajendra Mulay, Ms Mamata Joshi, Dr. Shraddha Chaudhary, Prof. Deepa Khushlani, and Prof. Malay Patra for their valuable support of this experiment.

## Supporting Information Available:

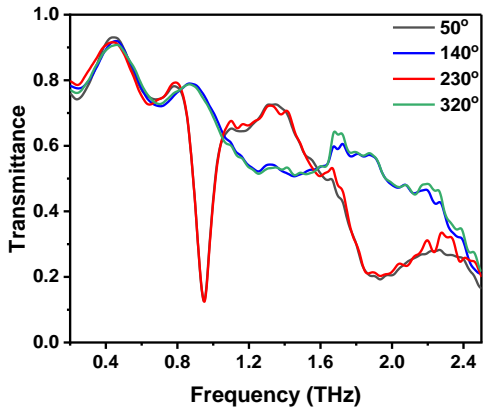
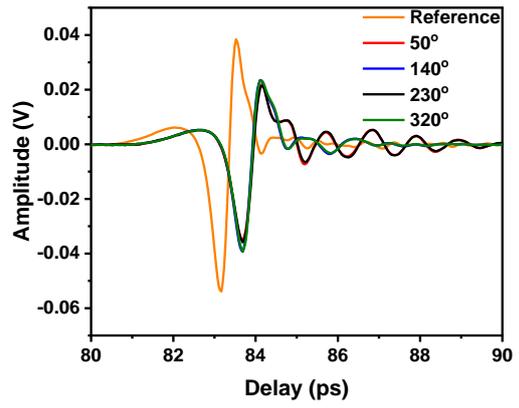
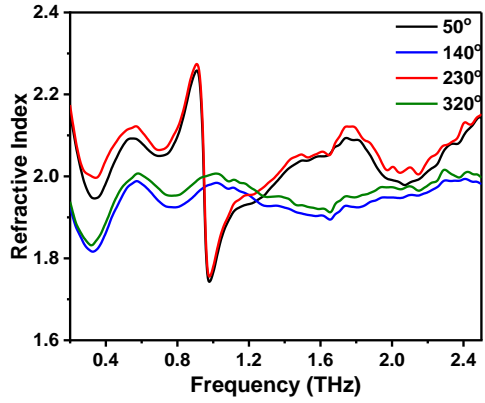
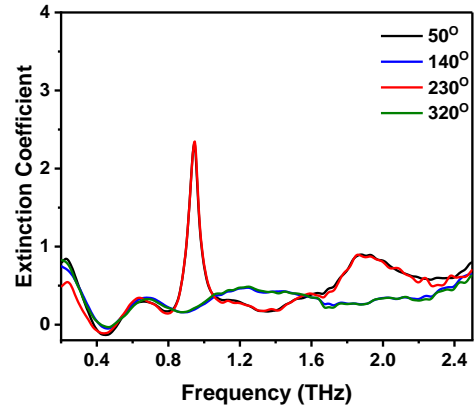
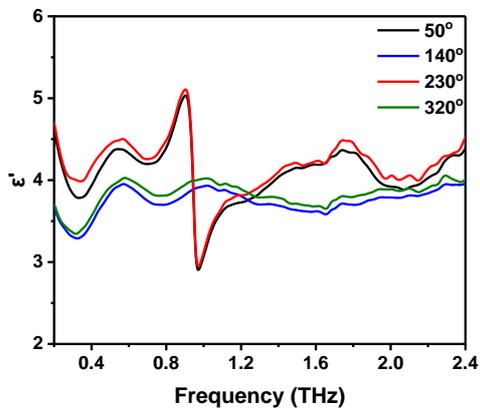
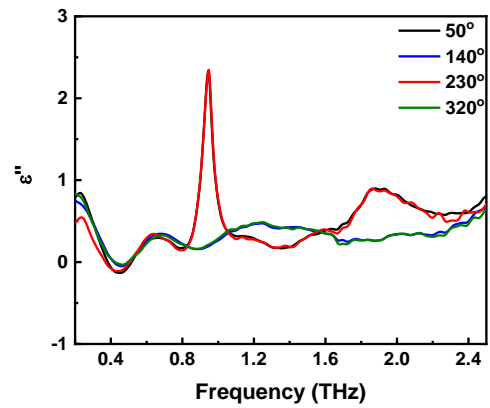
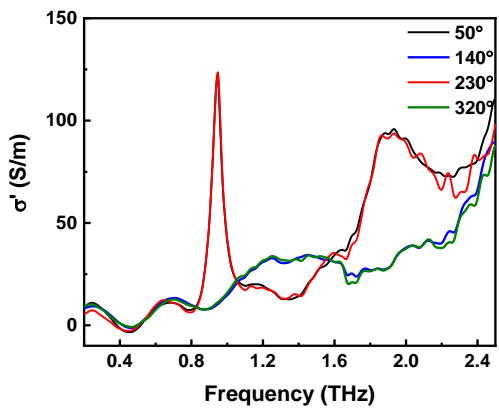
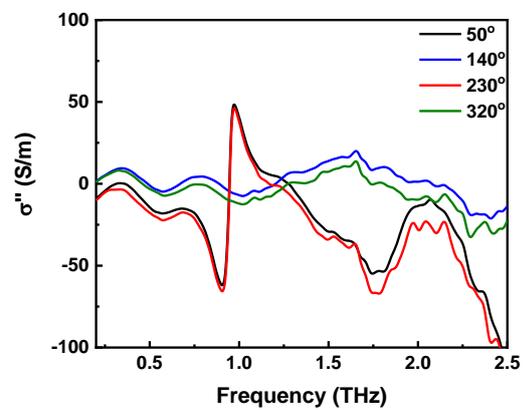